# Topologically-protected single-photon sources with topological slow light photonic crystal waveguides

Kazuhiro Kuruma,[1,2,3,*] Hironobu Yoshimi,[1,2] Yasutomo Ota,[4,5] Ryota Katsumi,[1,2] Masahiro Kakuda,[5] Yasuhiko Arakawa,[5] and Satoshi Iwamoto[1,2,5]

[1] *Research Center for Advanced Science and Technology, The University of Tokyo, 4-6-1 Komaba, Meguro-ku, Tokyo 153-8505, Japan*

[2] *Institute of Industrial Science, The University of Tokyo, 4-6-1 Komaba, Meguro-ku, Tokyo 153-8505, Japan*

[3] *John A. Paulson School of Engineering and Applied Sciences, Harvard University, Cambridge, MA 02138, USA*

[4] *Department of Applied Physics and Physico-Informatics, Keio University, 3-14-1 Hiyoshi, Kohoku-ku, Yokohama, Kanagawa 223-8522, Japan*

[5] *Institute for Nano Quantum Information Electronics, The University of Tokyo, 4-6-1 Komaba, Meguro-ku, Tokyo 153-8505, Japan*

*Corresponding author: kuruma@iis.u-tokyo.ac.jp

**Abstract**

**Slow light waveguides are advantageous for implementing high-performance single-photon sources required for scalable operation of integrated quantum photonic circuits (IQPCs), though such waveguides are known to suffer from propagation loss due to backscattering. A way to overcome the drawback is to use topological photonics, in which robust waveguiding in topologically-protected optical modes has recently been demonstrated. Here, we report single-photon sources using single quantum dots (QDs) embedded in topological slow light waveguides based on valley photonic crystals. We demonstrate Purcell-enhanced single-photon emission from a QD into a topological slow light mode with a group index over 20 and its robust propagation even under the presence of sharp bends. These results pave the way for the realization of robust and high-performance single-photon sources indispensable for IQPCs.**

# 1. Introduction

High-performance single-photon sources are recognized as a key element for scalable operation of integrated quantum photonic circuits (IQPCs) based on discrete variables. To implement such a quantum light source, it is essential to incorporate solid-state quantum emitters, such as semiconductor quantum dots (QDs), into photonic nanostructures for efficiently funneling their radiation into IQPCs. Among various platforms, slow light waveguides in photonic crystals (PhCs) [1] are particularly attractive. The slow light modes largely enhance light-matter interactions, which accelerate the emission of quantum emitters into the modes via the Purcell effect and thereby boost the performance of the single-photon source. In addition, the photonic bandgap effect in the PhCs strongly suppresses unwanted radiation into non-guided modes and thus further improves the source efficiency. Importantly, the slow light waveguides in PhCs have also been employed for studying nonlinear optics at single-photon levels [2], and chiral light-matter interactions [3]. However, PhC waveguides are known to often suffer from non-negligible propagation loss due to backscattering, which becomes more prominent when approaching the slow light regime [4,5]. Since the backscattering is induced by structural imperfections, which are inevitably accompanied with nanofabrication processes, a rather different approach might be necessary to solve the issue.

One possible strategy for mitigating the scattering loss in the slow light PhC waveguides is the use of topological photonics. Topologically-protected modes have been demonstrated to exhibit robust waveguiding immune to disorders [6]. Among reported topological waveguides, those based on valley photonic crystals (VPhCs) [7–14] have attracted much attention, since they can be realized using simple dielectric slabs. Therefore, VPhCs can be naturally implemented in conventional photonic integrated circuit platforms [7,8] and hence can be combined with QDs [9,11,15,16]. So far, the coupling of single QDs to topological waveguides have been demonstrated [17]. However, all these demonstrations utilized fast light modes that naturally emerge in topological bandgaps. In contrast to slow light modes, such fast light modes could exhibit only moderate enhancement in light-matter interactions.

In this Article, we report single-photon sources based on single QDs embedded in topological slow light VPhC waveguides. Single QDs coupled to the topological slow light waveguide exhibit large Purcell enhancement of spontaneous emission rate up to a factor of ~ 12. The generated single-photons robustly propagate through the slow light mode with a high group index ($n_g$) over 20 even with sharp waveguide bends. These results pave the way toward high-performance integrated quantum light sources required for IQPCs.

## 2. Device Structure

Figure 1a shows a schematic of the investigated topological VPhC waveguide with a bearded interface, which has been demonstrated to exhibit slow light modes in our recent work [18,19]. Compared to previously-reported topological slow light waveguides, which utilize complex materials or structures [20-26], the VPhC slow light waveguide can be realized in a simple dielectric slab and is advantageous for coupling to solid-state emitters. The waveguide is formed at the interface between two topologically-distinct VPhCs, denoted as type A (colored in blue) and type B (red). Figure 1b shows the unit cell of the VPhCs. The unit cell contains two equilateral triangle air holes with different side lengths termed $L_L$ and, $L_S$. In the case of $L_L = L_S$, $C_{6v}$ point group symmetry is preserved in the system and a symmetry protected Dirac cones are supported between the first and the second lowest frequency bands at K and K' points. On the other hand, the case of $L_L \neq L_S$ breaks spatial inversion symmetry of the system, resulting in formation of a topological bandgap between the two bands. Type A and B VPhCs coincide with each other when inverting one of them, and thus share the same band structure, but differ in their band topology. When interfacing the two VPhCs, a topological kink mode will appear within the common bandgap. Recent works have shown that the valley kink state can support robust light guiding even with sharp waveguide turns[7,8] and can significantly mitigate propagation loss and backscattering compared to conventional PhC waveguides [14,27].

Figure 1c shows a bearded interface formed using the two VPhCs with large and small triangles with $L_L =1.3a/\sqrt{3}$ and $L_S =0.7a/\sqrt{3}$ (*a* is lattice constant). Figure 1d shows a calculated band diagram of the VPhC waveguide with the bearded interface by the three-dimensional plane wave expansion (3D PWE) method. We consider a slab with the refractive index of $n = 3.4$ and thickness of $d = 130$ nm. We set the lattice constant *a* to be 310 nm. In stark contrast to conventional zigzag interfaces with fast light modes [9], the bearded interface used in this work supports two different slow light modes in the bandgap. They are degenerate at the Brillouin zone (BZ) edge due to glide plane symmetry across the interface [28]. The lower frequency mode corresponds to a topological edge state (light green curve), while the higher frequency one originates from a trivial state (black curve). The detailed discussion of the origin of these modes can be found in our previous work [18,19]. Figure 1d shows the corresponding group indices of the in-gap modes, showing high $n_g$ over 100 near the BZ edge.

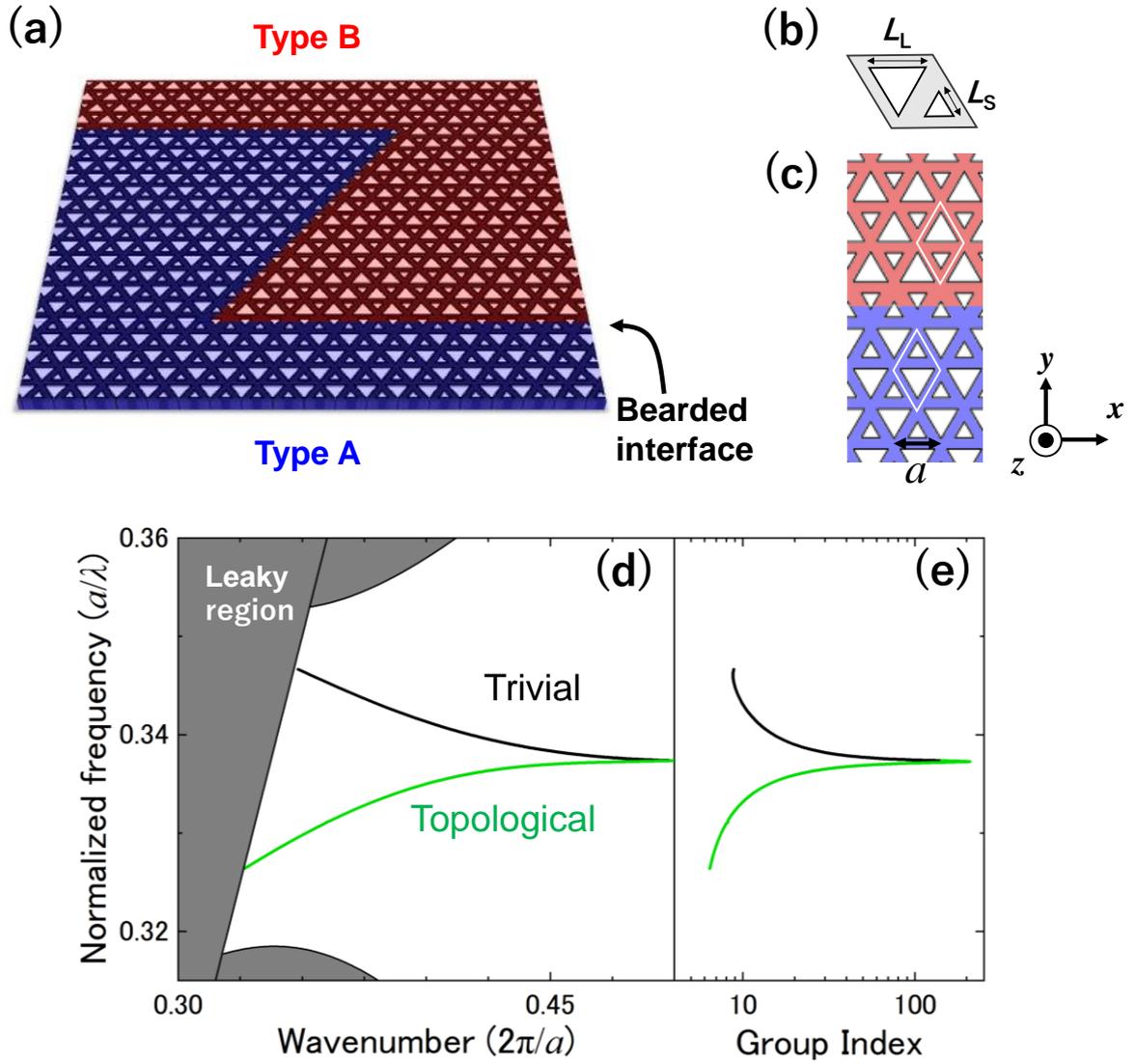

**Figure 1.** (a) Schematic of the investigated topological VPhC waveguide. (b) Unit cell of the VPhCs, consisting of large and small triangular air holes with respective side lengths of $L_L$ and $L_S$. (c) Bearded interface formed between two topologically-distinct VPhCs. The two unit cells in each VPhCs are indicated by white solid lines. (d) Band diagram for the edge state at the bearded interface. (e) Corresponding calculated group indices for the in-gap states.

## 3. Sample Fabrication and Basic Optical Characterization

We fabricated straight and Z-shaped topological slow light waveguides into a 130 nm-thick GaAs slab by standard semiconductor processes based on electron beam lithography and reactive ion etching. Air-bridge structures were finally formed by removing a 1 μm-thick $Al_{0.7}Ga_{0.3}As$ sacrificial layer under the slab using hydrofluoric acid. A single layer of InAs QDs with a low areal density of ~$10^8$ cm$^{-2}$ is contained in the middle of the slab. Photoluminescence (PL) peaks of individual QDs were observed in a spectral range from 900 to 1000 nm. Figures 2a, b show scanning electron microscope (SEM) images of one of the fabricated 52*a*-long straight and Z-shaped waveguides. Both ends of the waveguides are terminated with grating ports for light out-coupling to free space. We slightly shifted the position of the grating ports from the center of the waveguides to control the reflectance of light at the waveguide ends. For high reflective cases, we can observe Fabry-Pérot (FP) fringes in transmission spectra, enabling the extraction of $n_g$ of the waveguide modes [29].

First, we performed PL measurements to characterize the fabricated waveguides. The sample was placed in a liquid helium cryostat and pumped using a 775 nm pulse laser with a repetition rate of 80 MHz and a pulse duration of 1 ps. In order to clearly observe FP fringes in PL spectra, the sample temperature was kept at 60 K where confusing spectral peaks from single QDs are largely suppressed. We used a 50× objective lens to focus the laser light onto one grating port so as to excite the QDs embedded therein. The PL emission from the QDs was used as an internal light probe for measuring transmission spectra [9] and was collected from the other grating port via the same objective lens for analyzing with a spectrometer equipped with a Si CCD camera. Figure 2c shows the PL spectra of both straight and Z-shaped waveguides taken at a high average excitation power of 30 μW. We observed sharp peaks in the PL spectrum originating from FP resonances sustained by reflection at both waveguide ends. The FP fringes were observed from 910 nm to 1000 nm in the straight waveguide, while we did not see clear FP fringes in the shorter wavelength region below approximately 934 nm in the Z-shaped waveguide. The shorter wavelength band is the trivial mode with a high scattering loss when transmitting through the sharp waveguide bends. On the other hand, the longer wavelength band above 934 nm corresponds to the topological mode and thus can support robust light waveguiding even with the 120° bends. These interpretations for the observations are consistent with our previous results of numerical and experimental investigations on the VPhC slow light waveguides [18,19]. From the observed FP fringes, $n_g$ of both modes were estimated for the straight and Z-shaped waveguides (see Section S1, Supporting Information), and were plotted in Figure 2d. The measured $n_g$ for both waveguides show similar values in the topological band and rapidly increase near the wavelength of ~934 nm. The highest $n_g$ extracted from the measured FP fringes for the straight and Z-shaped waveguide are ~ 40 in the topological band. It is noted that we obtained higher $n_g$ of up to ~56 for topological modes in a different waveguide which exhibits a clearer FP spectrum particularly near its high $n_g$ region. These results demonstrate robust propagation of light emitted from QDs through topological slow light waveguides even under the presence of the sharp bends. The observed contrast between trivial and topological modes in PL spectra also highlights the impact of the topological protection in slow light waveguides. For the estimation of the BZ edge position of the fabricated topological waveguides, we fitted the measured $n_g$ for 11 different samples with the same design using theoretical curves computed by the 3D PWE method (solid lines in Figure 2d). From the fitting, we deduced the wavelength of the BZ edge to be 934.0 ± 0.45 nm. The fitting curve was plotted in Figure 2d and shows a good agreement with the measured $n_g$. Hereafter, using the extracted $n_g$ curve, we estimate $n_g$ of the waveguide mode only from its wavelength of operation, after taking into account a wavelength redshift of ~ 0.8 nm from 6.5K to 60K.

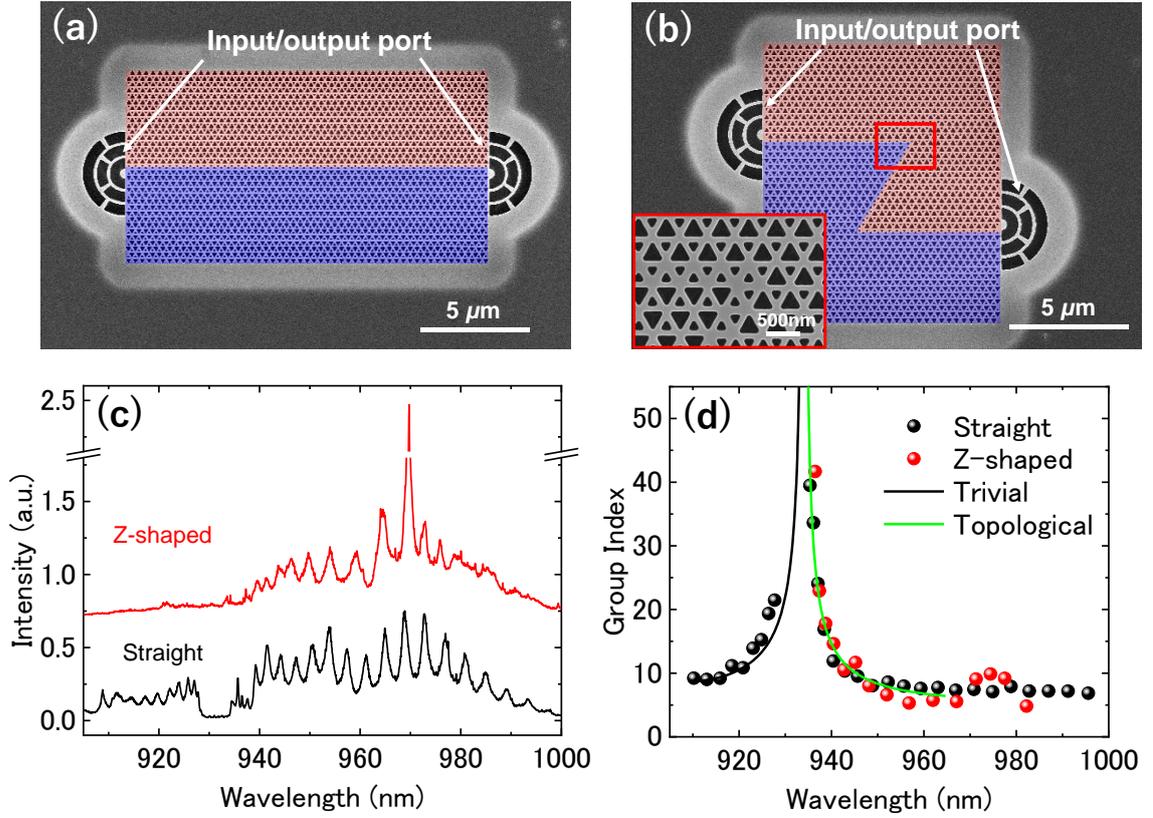

**Figure 2.** SEM images of fabricated (a) straight and (b) Z-shaped waveguides. The inset image in (b) shows an enlarged view of one of the 120° bends. The blue and red colored area corresponds to type A and type B VPhCs. (c) PL spectra of straight (black) and Z-shaped (red) waveguides. (d) Group indices extracted from the measured spectra in (c). The solid curves are simulated group indices of trivial (black) and topological (light green) modes.

## 4. Observation of Slow Light Enhanced Radiative Decay of Single QDs

Next, we optically investigated single QDs in straight waveguides at 6.5K. For inspecting each single QD, we excited a QD in a waveguide by impinging 775 nm laser pulses and measured its PL emission guided through the waveguide modes from the right-side output port one by one, as depicted in the inset of Figure 3a. Figure 3a shows an emission spectrum of a QD (labeled as QD-A) emitting at 936.17 nm. From the theoretical curve in Figure 2d, the emission wavelength can be translated into a high $n_g$ of $24 \pm 3$ in the topological waveguide band. We also performed time-resolved PL measurements on QD-A using a time-correlated single-photon counting system with a superconducting single-photon detector (SSPD). The measured system time resolution was 33 ps, which is dictated by the timing jitter of our SSPD. We selectively measured the QD emission using a spectrometer as a band-pass filter (bandwidth ~180 $\mu$eV). Figure 3b shows the time-resolved PL spectra measured for the QD-A (red) and a QD in an unpatterned area on the same chip (black). We fitted the measured PL curves using a single exponential function convolved with a peak function reflecting the measured system time resolution. The QD lifetime for QD-A is measured to be 0.6 ns, which is approximately 2 times shorter than the average lifetime for 10 QDs measured in unpatterned area (~ 1.2 ns). The observed lifetime reduction suggests that the emission rate of QD-A was enhanced by the Purcell effect in the topological slow light mode. Note that we also confirmed single-photon emission from the QD-A by intensity autocorrelation measurements.

We also investigated wavelength dependence of QD spontaneous emission rates for 49 different QDs in 10 different straight waveguides. Figure 3c shows the summary of the QD emission rates normalized by that of bulk QDs. Overall, the measured decay rates tend to increase as approaching the BZ edge of ~ 934 nm, which is consistent with the trend of measured $n_g$. We observed the highest enhancement of ~ 12 in the topological band (see Section S2, Supporting Information). We compared the measured decay rate with a theoretical model (see Section S3, Supporting Information), assuming the QD dipole moment is coupled to the $x$ component of the electric field dominant in the dielectric region (see Figure 3d). The theoretical decay rate was overlaid in Figure 3c with the same offset in the horizontal axis as that used for the theoretical curves in Figure 2d. The experimental data points with high decays agree well with the theoretical curves, further confirming the Purcell-enhanced emission of the QDs in the VPhC slow light modes. Meanwhile, many data points in Figure 3c were found below the theoretical curves. This observation indicates that the majority of the investigated QDs are located away from the field maxima of the waveguide modes and/or their polarization mismatched with those of the optical modes. This is quite common when using randomly-formed self-assembled QDs.

Figure 3d shows color maps of the Purcell factor ($F_p$) when a QD dipole moment is $x$- or $y$- linearly polarized ($d_x = d\hat{x}$ or $d_y = d\hat{y}$) and right- (subscript +) and left- (subscript −) circularly polarized ($d_\pm = d/\sqrt{2}(\hat{x}\pm i\hat{y})$, respectively). The $F_p$ map was calculated based on the model for a topological slow light mode with $n_g$ ~25. Whereas the maximum $F_p$ is found inside the air holes for both $E_x$ and $E_y$, there is still a certain amount of high $F_p$ in the dielectric region, especially for $E_x$ component. To achieve higher $F_p$, the precise alignment of the QD position and its polarization with respect to the local electric field could be necessary. Note that it seems difficult to simultaneously realize both good chiral and high Purcell-enhanced couplings of QDs to the current topological waveguide due to the mismatch between high chiral and high $F_p$ points in the dielectric region (see Section S4, Supporting Information), which, conversely, is suitable for chiral couplings with levitated atoms [30] and nanoparticles [31].

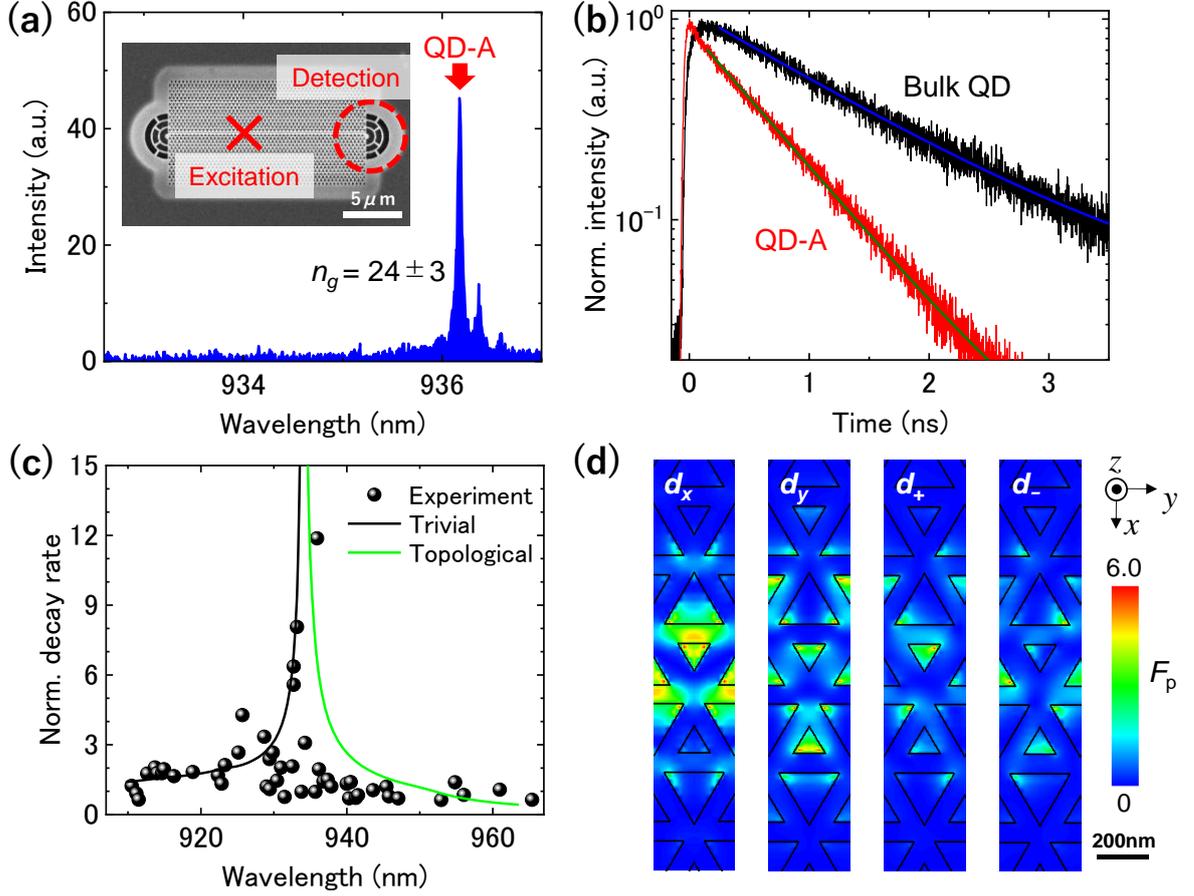

**Figure 3.** (a) PL spectrum for a QD (labeled as QD-A) in a straight waveguide. Inset shows the SEM image of a waveguide, overlaid with the positions of the laser excitation (cross marker) and QD emission detection (circle maker). (b) Measured PL decay curves of QD-A (red) and a bulk QD (black). The blue and green lines are fitting curves. (c) Normalized decay rates measured for 49 QDs. The black (light green) solid line indicates calculated decay rates for the electric field ($E_x$) of trivial (topological) modes. (d) Distribution of calculated Purcell factor ($F_p$) when a slow light mode with $n_g \sim 25$ is coupled to linear dipoles ($d_x$ or $d_y$) and circularly polarized dipoles ($d_+$ or $d_-$).

Finally, we investigate single-photon generation from a single QD in a Z-shaped waveguide.

As depicted in Figure 4a, we measured QD emission from the grating port located at the top left, while pumping a QD in the waveguide located far away from the grating port. Thus, we only detect PL signals that efficiently passed through the sharp corners. Figure 4b shows a PL spectrum of a QD (labeled QD-B) emitting at 936.03 nm (corresponding to $n_g$ of 25 ± 3). Figure 4c shows a measured time-resolved PL spectrum for QD-B, showing a large reduction in PL lifetime due to the Purcell effect by coupling to the slow light mode. By fitting the PL curves in the same procedure described above, the PL lifetime for the QD-B was deduced to be 0.34 ns, which is more than 3 times shorter than that of the bulk QD. We note that a higher Purcell enhancement of ~7 was also observed in a different Z-shaped waveguide (not shown).

In order to confirm the single-photon nature of the QD emission, we performed intensity autocorrelation measurements on QD-B using a Hanbury Brown-Twiss setup equipped with two SSPDs. Figure 4d shows a measured second-order correlation function, $g^{(2)}(t)$, measured using the output port. We observed a clear antibunching behavior with $g^{(2)}(0) = 0.26$. The non-zero value at $t = 0$ could be attributed to background emission from the FP fringes supplied by other QDs inside the waveguide. These results demonstrate Purcell-enhanced single-photon generation from a single QD in a topological slow light waveguide and its efficient propagation even under the presence of sharp bends.

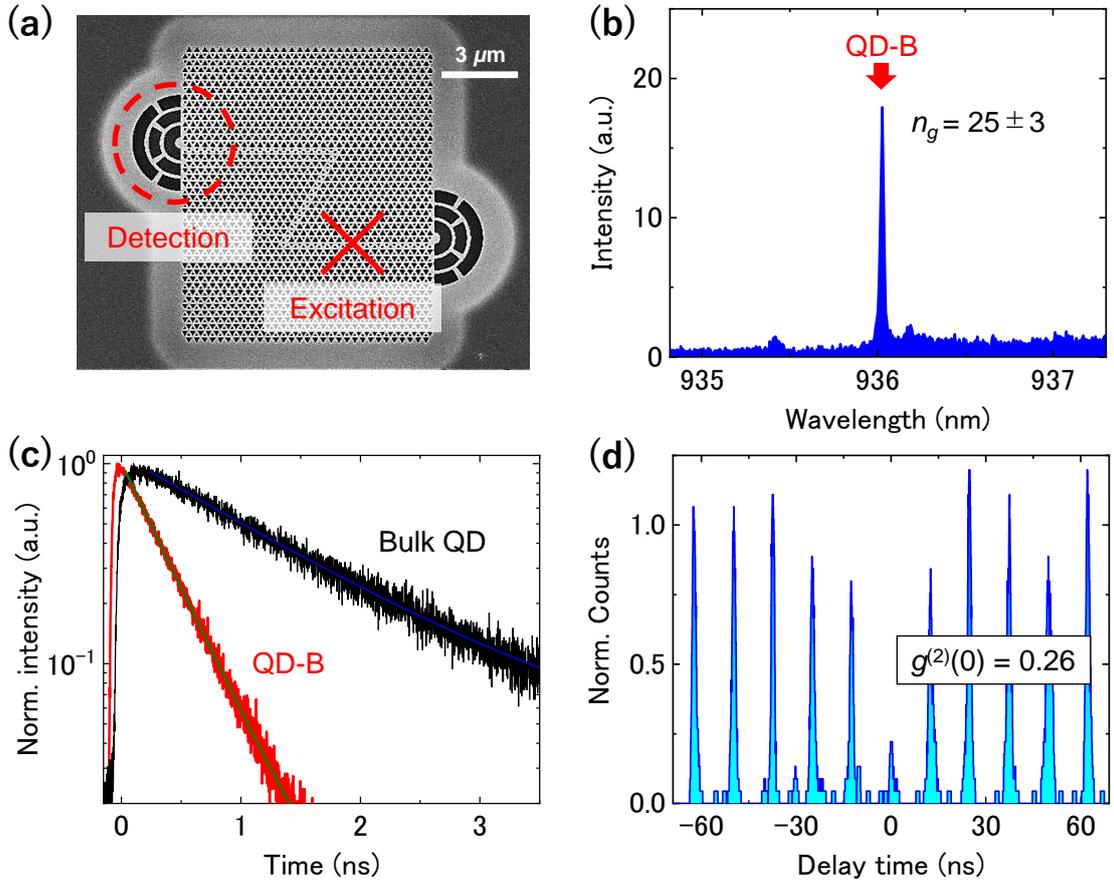

**Figure 4.** SEM image of a Z-shaped waveguide. The cross and circle makers indicate the positions of the excitation and detection. (b) PL spectrum of a QD (labeled as QD-B) measured at an averaged excitation power of 100 nW. (c) Time-resolved PL spectra measured for QD-B (red) and a bulk QD (black). (d) Measured second-order correlation function of QD-B.

## 5. Conclusion

In summary, we have demonstrated QD-based single-photon sources embedded in topological slow light VPhC waveguides. We observed a large reduction in PL lifetime of a single QD by a factor of up to ~12 due to the Purcell effect in the topological slow light modes. In the slow light regime with a high $n_g$ of ~25, we demonstrated Purcell-enhanced single-photon generation from a QD and its robust propagation even under the presence of sharp waveguide bends. These results will be of importance for the development of topologically-protected IQPCs with robust and high-performance single-photon sources.

**Acknowledgements**

We thank Prof. M. Lončar, M. Nishioka, Dr. S. Ishida, T. Yamaguchi and Dr. W. Lin for their technical support and helpful discussions. This work was supported by JSPS KAKENHI Grant-in-Aid for Specially Promoted Research (15H05700); KAKENHI (17H06138, 18J13565, 19K05300); JST-CREST (JPMJCR19T1); Asahi Glass Foundation; New Energy and Industrial Technology Development Organization (NEDO); JSPS Overseas Research Fellowships (202160592).

# Topologically-protected single-photon sources with topological slow light photonic crystal waveguides: supplemental document

Kazuhiro Kuruma,[1, 2, 3,*] Hironobu Yoshimi,[1,2] Yasutomo Ota,[4,5] Ryota Katsumi,[1,2] Masahiro Kakuda,[5] Yasuhiko Arakawa,[5] and Satoshi Iwamoto[1,2,5]

[1] *Research Center for Advanced Science and Technology, The University of Tokyo, 4-6-1 Komaba, Meguro-ku, Tokyo 153-8505, Japan*

[2] *Institute of Industrial Science, The University of Tokyo, 4-6-1 Komaba, Meguro-ku, Tokyo 153-8505, Japan*

[3] *John A. Paulson School of Engineering and Applied Sciences, Harvard University, Cambridge, MA 02138, USA*

[4] *Department of Applied Physics and Physico-Informatics, Keio University, 3-14-1 Hiyoshi, Kohoku-ku, Yokohama, Kanagawa 223-8522, Japan*

[5] *Institute for Nano Quantum Information Electronics, The University of Tokyo, 4-6-1 Komaba, Meguro-ku, Tokyo 153-8505, Japan*

*Corresponding author: kuruma@iis.u-tokyo.ac.jp

## S1. Estimation of group index ($n_g$)

From the observed Fabry-Pérot (FP) fringes in the transmission spectra in Figure 2c in the main text, we estimated $n_g$ for fabricated straight and Z shaped waveguides using the following equation:[1]

$$n_g = \frac{\lambda^2}{2L\Delta\lambda} \quad (S1)$$

where $\lambda$ and $\Delta\lambda$ are the center wavelength and the wavelength difference between two adjacent FP peaks, respectively. $L$ corresponds to the waveguide length of $52a$ ($a$ is lattice constant). To extract the wavelengths of FP peaks, we fitted the transmission spectrum with multiple Lorentzian peak functions. The data points in Figure 2d in the main text were derived using Equation S1.

**S2. Time-resolved PL spectrum of the sample with a Purcell enhancement of ~12**

Figure S1 shows a measured PL decay curve of a QD (labeled as QD-C) in a straight topological waveguide at 6.5K. The emission wavelength of QD-C is 935.92 nm (corresponding to $n_g$ =26 ± 3). The measured lifetime of QD-C was 98 ps, which is approximately 12 times shorter than that of the bulk QD (black curve).

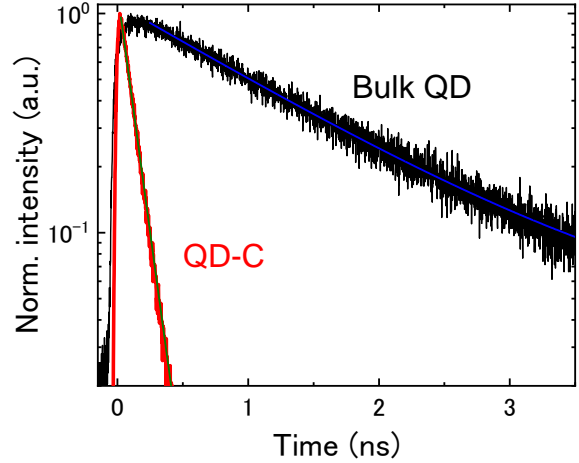

**Figure S1.**  Measured PL decay curves of QD-C (red) and a bulk QD (black). The blue and green lines are fitting curves.

**S3. Calculation of QD emission decay rates into waveguides**

In this calculation, we consider a GaAs-based two-dimensional photonic crystal waveguide shown in Figure 1c of the main text. The refractive index *n* of the GaAs slab is 3.4. The waveguide supports a transverse electric (TE) mode with a group index $n_g$. We assume QDs as point dipole sources (dipole approximation). The QD emission decay rate into a waveguide mode $\Gamma_{wg}$, normalized to the decay rate in bulk $\Gamma_0$ is given by[2]

$$\frac{\Gamma_{wg}}{\Gamma_0} = \frac{3}{4\pi} \frac{(\lambda/n)^2}{S_{eff}} \frac{n_g}{n} \frac{|E(r)\cdot d|^2}{|d|^2 |E_{max}|^2} \qquad (S2)$$

where $\lambda$ is the vacuum wavelength and $d$ is the electric dipole moment of the QD. The effective mode area is defined as $S_{eff} = \iint n(r)^2 |E(r)|^2 d^2r / \max[n(r)^2 |E(r)|^2]$. Here, $E(r)$ represents local electric field at a position $r$ in the waveguide. The electric field of the waveguide modes was calculated by the 3D plane wave expansion method. We assumed that the QD dipole is primarily coupled to the electric field towards the wave propagation direction ($E_x$). Since the QDs exist only in the GaAs slab, the maximum possible $\Gamma_{wg}$s were calculated using the maximum field intensity within the dielectric region. The calculated $\Gamma_{wg}$s for each waveguide mode plotted in Figure 3c in the main text have a constant offset of typical decay rate of InAs QDs in 2D photonic bandgap $\Gamma_{PC}$ ~ $0.1\Gamma_0$[3]. To obtain the profile of theorical Purcell factors ($F_p$) in the waveguide for $x$ or $y$-linear dipoles ($d_x$ or $d_y$), and right- or left- circularly polarized dipoles ($d_+$ or $d_-$), we calculated $F_p$ based on Equation (S2).

## S4. Mode map of Purcell factor ($F_p$) and chiral spatial map for the topological waveguide with a bearded interface

Figure S2a shows a calculated $F_p$ distribution for the case considering a right- or left-circularly polarized ($d_+$ or $d_-$) dipoles. $\sigma^+(\sigma^-)$ corresponds to the dipole moment of $d_+(d_-)$. A corresponding spatial map of normalized Stokes $S_3$ parameter, for a topological mode with a group index ($n_g$) of 25 is shown in Figure S2b. Normalized $S_3$ was calculated using the following equation:

$$S_3 = \frac{-2\,\text{Im}(E_x E_y^*)}{|E_x|^2 + |E_y|^2} \qquad (S3)$$

Here, $E_x$ and $E_y$ are the electric field of $x$ and $y$ components in waveguide. The locations at which $|S_3|$ approaches 1 will exhibit strong chiral coupling with a circularly polarized QD emission.

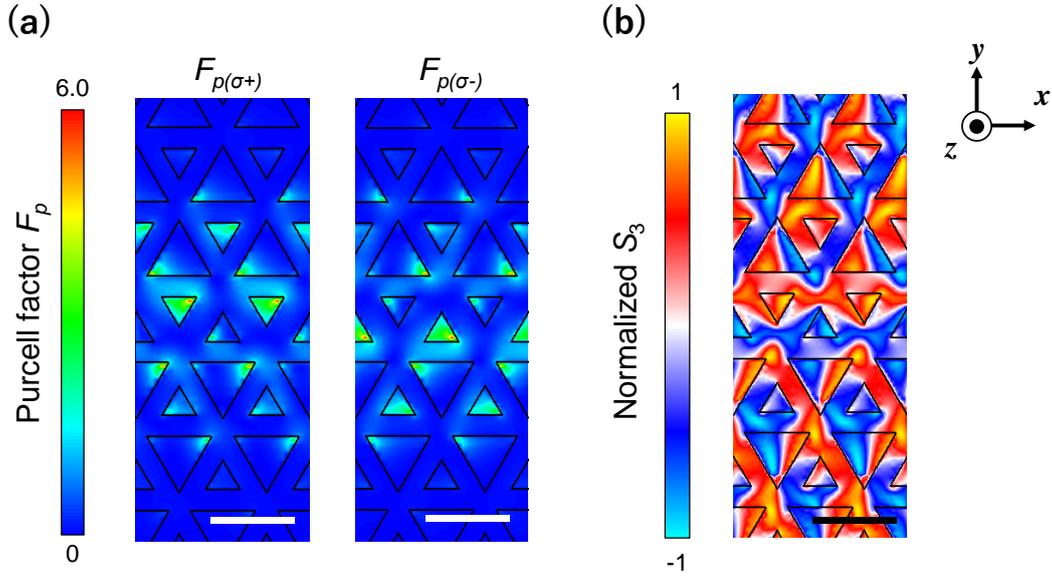

**Figure S2.** Mode profile of Purcell factor for σ⁺ and σ⁻ dipoles (a), and of normalized $S_3$ parameter (b) for a bearded topological waveguide with $n_g$ of 25. The white and black scale bars in (a) and (b) are 300 nm.